\documentclass[aps,pra,twocolumn,showpacs]{revtex4}
\usepackage{color}
\usepackage{graphicx}
\usepackage{amsmath}
\usepackage{amssymb}
\usepackage{amsfonts}
\usepackage{bm}
\def\epsbb{\mbox{\boldmath$\epsilon$}}
\def\epsbbs{\mbox{\boldmath\scriptsize$\epsilon$}}
\def\be{\begin{equation}}
\def\ee{\end{equation}}
\def\bea{\begin{eqnarray}}
\def\eea{\end{eqnarray}}
\def\rr{\mathbf{r}}
\def\qq{\mathbf{q}}
\def\RR{\mathbf{R}}

\begin{document}

\title{Heating rates for an atom in a far-detuned optical lattice}

\author{Fabrice Gerbier}
\affiliation{Laboratoire Kastler Brossel, \'Ecole Normale
Sup\'erieure, CNRS and UPMC, 24 rue Lhomond, 75231 Paris Cedex 05, France}

\author{Yvan Castin}
\affiliation{Laboratoire Kastler Brossel, \'Ecole Normale
Sup\'erieure, CNRS and UPMC, 24 rue Lhomond, 75231 Paris Cedex 05, France}

\begin{abstract}
We calculate single atom heating rates in a far detuned optical lattice, in connection with
recent experiments. We first derive a master equation, including a realistic atomic internal structure and
a quantum treatment of the atomic motion in the lattice. The experimental feature that optical lattices are
obtained by superimposing laser standing waves of different frequencies is also included, which leads to a micromotional
correction to the light shift that we evaluate. We then calculate, and compare to experimental results,
two heating rates, the ``total" heating rate
(corresponding to the increase of the total mechanical energy of the atom in the lattice), and
the ground bande heating rate (corresponding to the increase of energy within the ground energy band
of the lattice).
\end{abstract}

\date{\today}

\maketitle

\section{Introduction}

In the field of atomic quantum gases, optical lattices have become a versatile tool to trap atoms in an almost non-dissipative
way. This allows one to simulate many-body Hamiltonians originally formulated in
condensed matter physics, such as the Hubbard Hamiltonian for bosons and fermions (see \cite{bloch2008a} for
a recent review). This also opens promising implementations of quantum logical
operations \cite{mandel2003a}.
For all these applications, decoherence due to residual spontaneous emission of the atoms
in the optical lattice has to be kept as low as possible.
It was realized recently, on an experimental implementation of the Bose Hubbard model,
that a noticeable heating rate of the atoms takes place and has to be included
to obtain fair agreement with the theory \cite{trotzki2009a}.

In this article, we perform a theoretical analysis of the increase rate of the 
atomic mechanical energy in a far-detuned optical lattice. Contrarily to earlier laser cooling studies,
relying on semiclassical approximations \cite{Gordon1980,Cook,Kazantsev1981,dalibard1985a,bergsorensen1992,bergsorensen1994}, or restricting to the Lamb-Dicke
limit \cite{wineland1979a,cirac1992a}, or considering reduced dimensionalities and simplified level schemes
\cite{castin1991a,bergsorensen1993a,bergsorensen1994}, we include the relevant fine and hyperfine atomic structure and fully take into account
the quantum motion and tunneling of the atoms in the three-dimensional lattice. We however restrict for simplicity to the single atom problem:
We thus miss the photoassociation processes, which mainly produce losses of atoms \cite{xu2005a},
and the extra heating due to multiple scattering of fluorescence photons in the atomic gas.
Fortunately, as predicted in \cite{cirac1996,castin1998a} and observed experimentally in \cite{wolf2000a}, 
this extra heating is small in the so-called {\sl festina lente} limit, where
the fluorescence rate is much smaller than the oscillation frequency of an atom in the lattice,
a realistic regime for far-detuned optical lattices.

The article is organized as follows.
We present our model for the atomic structure and for the laser field producing the lattice, 
and we derive a master equation for the ground state
atomic density operator in section \ref{sec:model}.
We define and calculate two types of heating rates in sections \ref{sec:heating_total} and \ref{sec:heating_gb}.
In section \ref{sec:heating_total}
we show analytically that the total heating rate, defined
as the time derivative of the total mean mechanical energy of the atom,  remarkably is independent
of the atomic quantum state and insensitive to the sign, blue or red, of the laser detuning
with respect to the atomic resonance. 
In section \ref{sec:heating_gb},
we calculate the ground band heating rate,
that is the increase rate of the energy within the ground Bloch band of the lattice, for an atom
initially in the ground Bloch state; we show that
it widely depends on the blue or red nature of the lattice in the tight-binding limit.
We conclude in section \ref{sec:conclusion}.

\section{Model and ground state master equation}
\label{sec:model}

\subsection{Atomic structure}

\begin{figure}[t]
\centering{\includegraphics{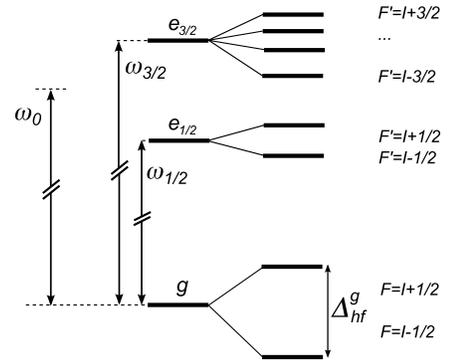}}
\caption{The atomic level scheme considered in this work, typical for alkali atoms.}
\label{fig:niveaux}
\end{figure}

We consider in this article alkali atoms, with an internal level structure shown schematically in Fig.~\ref{fig:niveaux}.
We consider only the ground $s$ state and the first $p$ excited states, and neglect the other excited states in our analysis. We denote by $g$ the ground state manifold, and by $e_{J}$, with $J=1/2$ or $3/2$ labelling the electronic angular momentum, the two excited fine multiplets 
(leading to the so-called $D_{1}$ and $D_{2}$ lines) separated in energy by an amount $\hbar\omega_{J}$ from the ground state. The fine structure atomic levels are further split by the hyperfine interactions $H_{\rm hf}$, giving rise to hyperfine multiplets with total angular momentum $F=I\pm1/2$ in the ground $s$ state (denoted by $g_{1}$ and $g_{2}$) and $F'=I\pm1/2$, $F'=I\pm3/2,I\pm1/2$ in the $e_{1/2}$ and $e_{3/2}$ manifolds, respectively. For order of magnitude estimates to come, we will note $\Delta_{\rm hf}^{g/e}$ typical values for the hyperfine splittings in the ground or excited states, respectively. The atomic density operator $\sigma$ has thus many internal atomic matrix elements, that may be collected in ground state elements, excited state elements and optical coherences, forming the submatrices
$\sigma_{gg} \equiv P_g \sigma P_g$, 
$\sigma_{e_{J} e_{J'}} \equiv P_{e_{J}} \sigma P_{e_{J'}}$,
and $\sigma_{e_{J} g} \equiv P_{e_{J}} \sigma P_{g}$.
Here $P_g$ projects onto the ground state manifold (including the hyperfine splitting in the two
ground states $g_{1}$ and $g_{2}$),  and $P_{e_J}$ projects onto the excited state fine multiplet $e_J$.

\subsection{Laser configuration}

In the experiment of \cite{trotzki2009a}, a three-dimensional cubic optical lattice was created by superimposing
standing waves along $x$, $y$, $z$, with orthogonal linear polarizations and with widely different frequencies.
The rapidly oscillating interference terms in the laser intensity between standing waves along
orthogonal directions average to zero and
the resulting light shift potential is essentially scalar.
We include this multichromatic feature in our model: Writing the total laser field as a sum of positive and negative frequency components, 
$\mathbf{E}(\rr,t)=\mathbf{E}^{(+)}(\rr,t) e^{-i\omega_{L}t}+\mathbf{E}^{(-)}(\rr,t) e^{i\omega_{L}t}$, with $\omega_L$ being the carrier
frequency, the amplitudes $\mathbf{E}^{(\pm)}(\rr,t)$ are taken as
\be
\label{eq:champ}
\mathbf{E}^{(+)}(\rr,t) = \sum_{\mu=x,y,z} \mathbf{e}_\mu \mathcal{E}_\mu(\rr) e^{-i\Delta_\mu t}
\ee
and $\mathbf{E}^{(-)}(\rr,t)=\left[\mathbf{E}^{(+)}(\rr,t)\right]^*$,
$\mathbf{e}_\mu$ being the unit vector along direction $\mu\in\{x,y,z\}$.
The modulation frequencies $\Delta_\mu$ associated with each axis $\mu=x,y,z$ are assumed to be incommensurate and much
smaller than the carrier frequency $\omega_L$.  We define the detunings from the excited states as
\begin{equation}
\delta_{J} = \omega_L - \omega_{J}
\end{equation}
where $\omega_{J}$ is the central resonance frequency of the atom on the $g\to e_{J}$
transition, see Fig.~\ref{fig:niveaux}. For simplicity, we assume that these detunings are much smaller than the atomic resonance frequency, so that the atom-laser coupling can be taken in the rotating wave approximation (RWA). This is consistent with our approximation where only the dominant $s-p$ transition
is considered. However, the detunings are assumed larger than any other frequency scale in the problem, including the residual modulation frequencies
\be
\label{eq:cond2}
\Delta_{\rm mod} \ll \vert \delta \vert.
\ee
Here and in what follows, when estimating orders of magnitude, we will write for simplicity $\delta$ as a typical order of magnitude for the detunings $\delta_{{J}}$ and $\Delta_{\rm mod}$ for the modulation frequencies $\Delta_{\mu}$.

In the rotating frame, the atom-laser coupling in the rotating wave approximation reads
\be
V_{\rm AL}(t) = -\mathbf{D}^{(+)}\cdot \mathbf{E}^{(+)}(\rr,t) + \mbox{h.c.}
\ee
where $\mathbf{D}^{(+)}$ is the raising part of the atomic dipole operator. 
The resulting time-averaged light shift potential is almost scalar, since, for the considered atomic structure,
the following ground state ``polarizability'' operator is scalar,
\be
\label{eq:scalarite}
(\mathbf{D}\cdot\mathbf{e}_L)_{g e_{J}} (\mathbf{D}\cdot\mathbf{e}_L)_{e_{J} g}=d_{{J}}^2 P_g
\ee
where $\mathbf{e}_L$ is any unit vector with real components,
and $(X)_{e_{J} g}=P_{e_{{J}}} X P_{g}$ is the restriction of any operator $X$ between the manifolds
$e_{J}$ and $g$. Eq.~(\ref{eq:scalarite}) may be checked in the fine structure basis from the expression of the Clebsch-Gordan coefficients,
where it appears as a well known property of $1/2\to 1/2$ and $1/2\to 3/2$ transitions. From the orbital rotational invariance of $(\mathbf{D})_{eg}
\cdot (\mathbf{D})_{ge}= d^2 P_e$, one also deduces the relation
\begin{equation}
\label{eq:def_da}
\left(\mathbf{D}\right)_{e_{J} g} \cdot \left(\mathbf{D}\right)_{g e_{J'}} =
d^2 P_{e_{J}} ~\delta_{{J}{J'}} 
\end{equation}
where $d$ is the so-called reduced atomic dipole moment for the $s-p$ transition. This finally leads to
\be
d_{1/2}^2 = \frac{d^2}{3} \ \ \ \mbox{and}\ \ \ 
d_{3/2}^2 = \frac{2d^2}{3}.
\ee

\subsection {Equations of motion} 
\label{subsec:eom}

The starting point of our treatment is the fully quantum optical Bloch equation for the atomic density operator $\sigma$
\cite{cct1992a}, including
the hyperfine structure and 
a quantum treatment of the external atomic variables (center of mass position).
We assume the following, typical hierarchy between the different energy scales in the problem:
\be
\label{eq:cond_deplct}
E_{\rm rec}, E_{\rm kin}, U_{\rm max}  \ll \hbar \Gamma \ll \vert \hbar \delta \vert.
\ee
Here $E_{\rm rec}$ is the atomic recoil energy,
\be
\label{eq:erec}
E_{\rm rec}=\frac{\hbar^2 k_L^2}{2m},
\ee
$k_L=\omega_L/c$ is the laser wavevector, 
$\Gamma$ is the natural linewidth of the excited states,
$E_{\rm kin}$ is the  atomic kinetic energy, and $U_{\rm max}\sim \vert \hbar\Omega^2/\delta \vert$ 
is the typical depth of the optical lattice potential, with a laser induced Rabi frequency $\Omega$ loosely defined by $V_{\rm AL}\approx \hbar \Omega$. The condition $U_{\rm max} \ll \hbar |\delta|$ implies the weak saturation limit, $\frac{\Omega^2}{\delta^2} \ll 1$. In these conditions, we can adiabatically eliminate the fast optical coherences and excited state matrix elements, which track the slowly-evolving ground state variables. The large detuning regime Eqs.(\ref{eq:cond2},\ref{eq:cond_deplct}) considerably simplifies this elimination, allowing one to perform an expansion of $\sigma_{eg}$ and $\sigma_{ee}$
in powers of $1/\delta$, up to order $1/\delta^2$ here.

As detailed in the Appendix \ref{app:deriv},
this treatment leads to an effective equation for the ground state density matrix $\sigma_{gg}$, which is still rather complicated due to the hyperfine Hamiltonian and to the explicit time-dependence introduced by the temporal modulation of the laser electric field at frequencies $\Delta_\mu$. The light shifts in particular have a non-scalar part, that also provides a Raman coupling between $g_1$ and $g_2$  with a Rabi frequency $\approx \Omega^2/\delta$. The equations can be further simplified in the experimentally relevant case, where the modulation frequency is much faster than the atomic motion in the optical lattice, but much smaller than the ground state hyperfine splitting (so as to make the hyperfine
Raman coupling widely non-resonant),
\be
\label{eq:cond_dmod}
\frac{\Omega^2}{\delta} \ll \Delta_{\rm mod} \ll \Delta_{\rm hf}^g.
\ee
The inequalities in Eq.~(\ref{eq:cond_dmod}) have two consequences. 

First, due to the smallness of the laser induced Raman couplings as compared to the ground state hyperfine splitting, 
ground state hyperfine coherences
will depart from their initial zero value by a small amount, of order
$\Omega^2/(\delta \Delta_{\rm hf}^g)$. Their adiabatic elimination (valid in the absence of Raman resonances) leads to
a small correction to the light shift of order
\be
\label{eq:du1}
\delta U_{\rm hf} \approx \frac{\hbar \Omega^4}{\delta^2 \Delta_{\rm hf}^g}.
\ee
After this second adiabatic elimination, the unknown part of the density operator is now the diagonal part 
$\sigma_{gg}^{\rm diag}$ of $\sigma_{gg}$, the diagonal part of any ground state operator $X$
being defined as
\be
X^{\rm diag} = P_{g_1} X P_{g_1} +  P_{g_2} X P_{g_2},
\ee
where $P_{g_i}$ projects onto the ground state $g_i$, $i\in\{1,2\}$. Of course, $\sigma_{gg}^{\rm diag}$ still contains Zeeman coherences within the $g_1$ and 
$g_2$ manifolds.

Second, due to the time-dependence of the laser field at frequencies $\Delta_{\mu}$, $\sigma_{gg}^{\rm diag}$ can be decomposed as a slowly-evolving, zero frequency component $\bar{\sigma}^{\rm diag}_{gg}$ plus fast oscillating components at harmonics of the modulation frequencies $\Delta_{\mu}$. The latter correspond to an atomic micromotion in the time-dependent optical lattice, similar to the dynamics in Paul-type ion traps \cite{paul1990a}. The fast components are typically much smaller than the slow one by a factor $\approx \Omega^2/(\delta \Delta_{\rm mod})$. In this regime, a perturbative technique \cite{rahav2003a,ridinger2009a} allows to construct a time-independent effective potential induced on $\bar{\sigma}_{gg}^{\rm diag}$ by this micromotion, which constitutes another small correction to the light shift,
explicitly calculated in the Appendix \ref{app:deriv}, and of order of magnitude
\be
\label{eq:du2}
\delta U_{\rm micro} \approx E_{\rm rec} \frac{\Omega^4}{\delta^2 \Delta_{\rm mod}^2}
\ee
If $\Delta_{\rm mod}^2\gg E_{\rm rec}\Delta_{\rm hf}^g$, as is the case in \cite{trotzki2009a}, the hyperfine contribution in Eq.~(\ref{eq:du1}) dominates over the  micromotion contribution in Eq.~(\ref{eq:du2}). 

The hyperfine (\ref{eq:du1}) and micromotion (\ref{eq:du2}) contributions appear at order $\delta^{-1}$ of the adiabatic elimination 
of $\sigma_{ee}$ and $\sigma_{eg}$, and
originate from a purely conservative (though time-dependent) light shift potential.
At order $\delta^{-2}$ in the adiabatic elimination,
one obtains non-conservative terms, proportional to the fluorescence rate
$\Gamma (\Omega/\delta)^2$, and corrections to the light shift potential of order
\be
\label{eq:du3}
\delta U_{1/\delta^2} \approx \left(\frac{\Omega}{\delta}\right)^2  \hbar \Delta_{\rm hf}^{e,g}, 
\left(\frac{\Omega}{\delta}\right)^2\hbar \Delta_{\rm mod}.
\ee
All these terms may be directly time averaged, neglecting micromotion corrections at this order. 

This procedure results in the final master equation for the zero frequency hyperfine diagonal part of the ground state density operator,
\begin{multline}
\label{eq:ggf}
\frac{d}{dt} \bar{\sigma}_{gg}^{\rm diag} = \frac{1}{i\hbar} [\frac{\mathbf{p}^2}{2m}+ U + \delta U,
\bar{\sigma}_{gg}^{\rm diag}] -\frac{1}{2} \{W, \bar{\sigma}_{gg}^{\rm diag}\} \\
+\left(\mathcal{L}[\bar{\sigma}_{ee}^{(2)}]\right)^{\rm diag}.
\end{multline}
The structure of this equation is familiar from earlier studies on laser cooling. The first commutator corresponds to a Hamiltonian evolution in the light shift potential $U+\delta U$. For our choice of polarizations and detunings, the basic scalar light shift potential $U$ is scalar,
\be
\label{eq:basic}
U(\rr) = \left(\sum_{J} \frac{d_{{J}}^2}{\hbar\delta_{J}}\right) P_g 
\left(\sum_\mu |\mathcal{E}_\mu(\rr)|^2\right).
\ee
The quantity $\delta U$, whose complete expression is given in the Appendix \ref{app:deriv}, includes all previously discussed corrections in Eq.~(\ref{eq:du1},\ref{eq:du2},\ref{eq:du3}).
As discussed in the caption of Table~\ref{Table1}, all these corrections are negligible to a good accuracy for typical experimental parameters.

The part in Eq.~(\ref{eq:ggf}) involving an anti-commutator with the operator $W$ corresponds formally to an anti-hermitian Hamiltonian, and describes departure from the ground state upon absorption of a laser photon. This part is also
scalar as expected,
\be
W(\rr) = \left(\sum_{J} \Gamma \frac{d_{{J}}^2}{(\hbar\delta_{J})^2} \right) P_g
\left(\sum_\mu |\mathcal{E}_\mu(\rr)|^2\right).
\ee
Finally, the last ``feeding'' term in Eq.~(\ref{eq:ggf}) describes atoms returning to the ground state after an absorption-spontaneous emission cycle. 
This part involves a Liouvillian operator $\mathcal{L}$ acting on the excited state density operator,
\begin{multline}
\label{eq:feed}
\mathcal{L}[\sigma_{ee}]=
\frac{3\Gamma}{8\pi d^2}  \sum_{{J},{J'}} \int d^2n \sum_{\epsbbs\perp\mathbf{n}}
\\
e^{-ik_0 \mathbf{n}\cdot\mathbf{r}}
\left(\mathbf{D}^{(-)}\cdot\epsbb^*\right)_{g e_{J}}
\sigma_{e_{J} e_{J'}}
\left(\mathbf{D}^{(+)}\cdot\epsbb\right)_{e_{J'} g}
e^{ik_0 \mathbf{n}\cdot\mathbf{r}}
\end{multline}
where $\Gamma = d^2k_0^3/(3\pi\epsilon_0\hbar)$ is the usual spontaneous linewidth, and where
the fine energy splitting in the momentum $k_0$ of the spontaneously emitted photon was neglected.
In (\ref{eq:ggf})  $\mathcal{L}$ acts on the zero frequency component of the excited state density operator expressed in terms of $\bar{\sigma}_{gg}$ as
\be
\label{eq:eeb}
\bar{\sigma}_{e_{J} e_{J'}} = 
\frac{1}{\hbar^2\delta_{J}\delta_{J'}} 
\sum_\mu (D_\mu)_{e_{J} g} \mathcal{E}_{\mu}
\bar{\sigma}_{gg}^{\rm diag} (D_\mu)_{g e_{J'}} \mathcal{E}_{\mu}^*.
\ee
If one takes the trace over the internal atomic states of this expression, in the case ${J}={J'}$,
one obtains from Eq.~(\ref{eq:scalarite}) the useful property
\be
\label{eq:utile}
\mbox{Tr}_{\rm int}(\bar{\sigma}_{e_{J} e_{J}}) =
\frac{d_{{J}}^2}{(\hbar\delta_{J})^2} 
\sum_\mu \mathcal{E}_\mu(\rr) \mbox{Tr}_{\rm int}(\bar{\sigma}_{gg}^{\rm diag})
\mathcal{E}_\mu^*(\rr).
\ee

\section{Total heating rate}
\label{sec:heating_total}

From the master equation Eq.~(\ref{eq:ggf}), we now calculate the rate of change $dE_{\rm tot}/dt$ of the total mean mechanical energy
of the atom:
\be
E_{\rm tot} = \mbox{Tr} [(\frac{\mathbf{p}^2}{2m} + U+\delta U) \bar{\sigma}_{gg}^{\rm diag}],
\ee
where averaging is done on both internal and external degrees of freedom. The first commutator in the right hand side of Eq.~(\ref{eq:ggf}) 
does not contribute to $dE_{\rm tot}/dt$. For the other two terms in the right hand
side of Eq.~(\ref{eq:ggf}), we calculate exactly the contribution
involving $\mathbf{p}^2/2m$ and $U$ in the energy, and we put a simple bound on the contribution
involving  $\delta U$.
Since $\mathbf{p}^2/2m$ and $U$ are scalar, all internal traces may be evaluated,
and we obtain
\begin{multline}
\label{eq:de1}
\frac{d}{dt} E_{\rm tot} = 
\left(\sum_{{J}=\frac{1}{2},\frac{3}{2}}
\Gamma\frac{d_{{J}}^2}{(\hbar\delta_{J})^2}\right)
\sum_{\mu=x,y,z} 
 \Big\langle \frac{\hbar^2 k_{0}^2}{2m} |\mathcal{E}_\mu(\rr)|^2  \\
+\mathcal{E}_\mu^*(\rr) \frac{\mathbf{p}^2}{2m} \mathcal{E}_\mu(\rr) 
-\frac{1}{2} \{ \frac{\mathbf{p}^2}{2m}, |\mathcal{E}_\mu(\rr)|^2\}\Big\rangle \\
+O[\Gamma \left(\Omega/\delta\right)^2 \delta U],
\end{multline}
where $\langle X\rangle=\mbox{Tr}[X\bar{\sigma}_{gg}^{\rm diag}]$ is the expectation
value of any ground state observable $X$.
To obtain this result, we have taken the sum over $\epsbb$ 
and the integral over $\mathbf{n}$ in Eq.~(\ref{eq:feed}), using
$e^{i k_0 \mathbf{n}\cdot\mathbf{r}} H_0
e^{-i k_0 \mathbf{n}\cdot\mathbf{r}} = H_0 -\mathbf{p}\cdot
\hbar k_0 \mathbf{n}/m + \hbar^2 k_0^2/(2m)$, where $H_0= \mathbf{p}^2/(2m)+U+\delta U$; the term linear
in $\mathbf{n}$ is odd and vanishes after integration over $\mathbf{n}$.
We also used Eq.~(\ref{eq:def_da}) and Eq.~(\ref{eq:utile}).

The last term in Eq.~(\ref{eq:de1}), originating from $\delta U$, is negligible
as compared to first term, originating from the recoil due to spontaneous emission,
provided that 
\be
\label{eq:conddU}
|\delta U|\ll \frac{\hbar^2 k_0^2}{2m}.
\ee
Using the estimates given in Eqs.(\ref{eq:du1},\ref{eq:du2},\ref{eq:du3})
we find that this is extremely well obeyed in \cite{trotzki2009a}.
Condition Eq.~(\ref{eq:conddU}) is supposed in what follows to be satisfied. 

Eq.~(\ref{eq:de1}) can be further transformed,
using the fact that Maxwell's equation imposes 
$(\Delta_{\rr} + k_L^2) \mathcal{E}_\mu(\rr) =0$
where $k_L=\omega_L/c$ and $\Delta_\rr$ is the Laplacian operator.
Then
\begin{multline}
\label{eq:de2}
\frac{d}{dt} E_{\rm tot} = \left(\sum_{{J}} \Gamma \frac{d_{{J}}^2}{(\hbar\delta_{J})^2}\right)
\\ \times
\sum_\mu 
\Big\langle
\frac{\hbar^2 k_0^2}{2m} |\mathcal{E}_\mu(\rr)|^2
+\frac{\hbar^2}{2m} |\mathrm{grad}\,\mathcal{E}_\mu(\rr)|^2
+\mathbf{j}_\mu\cdot\mathbf{p}
\Big\rangle
\end{multline}
where the laser field current for the field amplitude $\mathcal{E}_\mu$
is $\mathbf{j}_\mu=\hbar(\mathcal{E}_\mu^*\mathrm{grad}\,\mathcal{E}_\mu
-\mbox{c.c.})/(2im)$ and one also  has 
$|\mathrm{grad}\,\mathcal{E}_\mu|^2=k_L^2|\mathcal{E}_\mu|^2+\Delta_{\rr} |\mathcal{E}_\mu|^2/2$.
Furthermore, if each amplitude $\mathcal{E}_\mu$ is a simple laser standing wave, $\mathcal{E}_\mu(\rr)=
\mathcal{E}_{0,\mu} \cos(\mathbf{k}_\mu\cdot\mathbf{r}+\varphi_\mu)$,
and under the reasonable assumption that $k_0$ may be identified to the laser wavevector $k_L$,
one finally obtains
\be
\label{eq:de3}
\frac{d}{dt} E_{\rm tot} = \frac{\hbar^2k_L^2}{2m} \left(\sum_{J} 
\Gamma \frac{d_{{J}}^2}{(\hbar\delta_{J})^2}\right)
\left(\sum_\mu|\mathcal{E}_{0,\mu}|^2\right).
\ee
This may be rewritten in terms of the maximal fluorescence rate $\Gamma_{\rm fluo}^{\rm max}$
in the lattice, that is the maximum of $W$:
\be
\frac{d}{dt} E_{\rm tot} = \frac{\hbar^2k_L^2}{2m} \Gamma_{\rm fluo}^{\rm max} .
\label{eq:forme_utile}
\ee
In this form, (\ref{eq:forme_utile})  
reproduces in the classical limit the position-independent value of the momentum diffusion of a two-level atom in a weak laser standing wave,
obtained from laser cooling theory \footnote{In the semi-classical treatment, one writes a Fokker-Planck
equation for the density in phase space \cite{dalibard1985a}. Taking the zero velocity limit $v=0$
of the mean force and diffusion tensor $D$, as allowed by the condition
$kv\ll \Gamma \ll |\delta|$, one finds $dE_{\rm tot}/dt= \mbox{Tr} D/m$. An explicit calculation
of the diffusion coefficient was performed for a two-level atom \cite{Cook,Gordon1980}. In the present case of
large detuning, the light shift is scalar and the underlying $1/2\to 1/2$ and
$1/2\to 3/2$ transitions may be modeled by a two-level atom, up to a global normalisation
of the fluorescence rate.}.

Eq.~(\ref{eq:de3}) is one of our main results. It shows the counterintuitive fact 
that for a given laser intensity distribution, the total atomic heating rate in a far-detuned optical lattice is independent of the atomic external state and of the sign of the atom-laser detuning, provided that each laser standing wave is linearly polarized
and the atomic kinetic energy remains small as compared to $\hbar\Gamma$.
In particular, trapping the atoms at the nodes of a blue detuned optical lattice,
although it reduces the atomic fluorescence rate, does not reduce the total heating rate $dE_{\rm tot}/dt$ with respect to trapping at the antinodes of a red detuned optical lattice with the same absolute value of the detuning and laser
intensity.

\section{Ground band heating rate}
\label{sec:heating_gb}

\begin{figure}[t]
\centering{\includegraphics{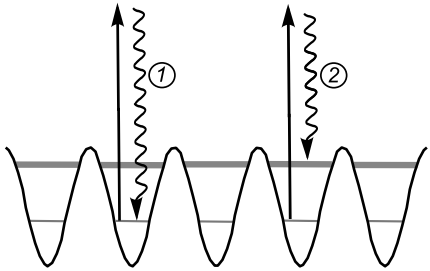}}
\caption{Sketch of the different heating processes in a lattice in the tight-binding limit: intraband
processes (1) and interband processes (2).}
\label{fig:ReseauES}
\end{figure}

We now consider the increase rate of energy within the ground band of the optical lattice. 
For experiments with atomic gases, the physical motivation to do so is twofold: One can imagine observable quantities
that depend mainly on the probability amplitudes to find the atoms
in the Bloch states of the ground band, and one can also imagine that evaporative-type
experimental techniques are developed to continuously eliminate the atoms from the excited energy bands.

Having previously bounded the effect of the light shift correction $\delta U$, we directly neglect it in the
master equation Eq.~(\ref{eq:ggf}), and we define the ground band mean energy as
\be
E_{\rm GB} = \frac{\mbox{Tr}[P_{\rm GB} (\frac{\mathbf{p}^2}{2m} + U) \bar{\sigma}_{gg}^{\rm diag}]}
{\mbox{Tr}[P_{\rm GB}\bar{\sigma}_{gg}^{\rm diag}]}
\ee
where $P_{\rm BG}$ projects onto the ground energy band in the periodic potential $U$.
The increase rate of $E_{\rm GB}$, contrarily to the total energy increase rate, depends on the atomic state.
To make the analytical calculation tractable \footnote{Numerically, one can go beyond this assumption
with the Monte Carlo wavefunction technique used in \cite{Castin3D} to study
three-dimensional laser cooling.}, we thus restrict to the initial
increase rate $\frac{d}{dt}E_{\rm GB}(t=0)$ for an initial atomic state only
populating the ground Bloch state, that is the state with zero Bloch vector
$\qq=\mathbf{0}$ in the ground band (of band indices $\mathbf{0}$). This choice is motivated
by the fact that the heating rate of \cite{trotzki2009a}
was measured for a Bose-Einstein condensate.

We shall present numerical results in the general case, and then
focus on the tight-binding limit, with a modulation depth of the optical potential $U(\rr)$ much
larger than the atomic recoil energy, where analytical results are obtained.
It is then convenient to assume that the optical lattice potential has a minimum at the origin $\mathbf{r}=\mathbf{0}$
of the coordinates. We thus take 
$\mathcal{E}_\mu = \mathcal{E}_{0,\mu} \cos(\mathbf{k}_\mu\cdot\mathbf{r})$
for a red detuned lattice ($\delta_{J} <0$ $\forall {J}$),
and 
$\mathcal{E}_\mu = \mathcal{E}_{0,\mu} \sin(\mathbf{k}_\mu\cdot\mathbf{r})$
for a blue detuned lattice ($\delta_{J} >0$ $\forall {J}$). 

\noindent {\sl General results:} 
The initial atomic density operator is:
\be
\bar{\sigma}_{gg}^{\rm diag}(t=0) = |\mathbf{0};\qq=\mathbf{0}\rangle \langle \mathbf{0};\qq=\mathbf{0}| 
\otimes \sigma_{gg}^{\rm int}.
\ee
Here $\sigma_{gg}^{\rm int}$ is any ground internal atomic state (with no hyperfine coherences),
$|\mathbf{0};\mathbf{q}\rangle$ is the ground band eigenstate of Bloch vector $\qq$,
normalized in an arbitrarily
large quantization volume commensurate with the lattice spacing, 
and (for futur reference) $E_{\mathbf{0};\mathbf{q}}$ is the corresponding eigenenergy.
Since $\bar{\sigma}_{gg}^{\rm diag}(t=0)$ commutes with $H_0=\mathbf{p}^2/(2m)+U(\rr)$ 
and with the projector $P_{\rm GB}$, the time
derivative of the ground band energy at $t=0$ has, from (\ref{eq:ggf}), the simple expression
involving the feeding term only:
\be
\frac{d E_{\rm GB}}{dt}(t=0) = \mbox{Tr}[P_{\rm GB} (H_0-E_{\mathbf{0};\qq=\mathbf{0}})
\mathcal{L}[\bar{\sigma}_{ee}(0)].
\ee
Replacing the feeding operator by (\ref{eq:feed}) and the excited state density operator
by its expression (\ref{eq:eeb}), we obtain at time $t=0$:
\begin{multline}
\frac{d E_{\rm GB}}{dt}= 
\Gamma \left(\sum_J \frac{d_J^2}{(\hbar\delta_J)^2} \right)
\int \frac{d^2n}{4\pi} \sum_\mu \left[1+\left(\frac{1}{3}-n_\mu^2\right)C\right] \\
\times (E_{\mathbf{0};k_L(\mathbf{e}_\mu-\mathbf{n})}- E_{\mathbf{0};\mathbf{0}})
|\langle \mathbf{0}; k_L(\mathbf{e}_\mu-\mathbf{n}) |
e^{-ik_L \mathbf{n}\cdot\mathbf{r}} \mathcal{E}_\mu(\mathbf{r}) |\mathbf{0};\mathbf{0}\rangle|^2.
\label{eq:degb_gen}
\end{multline}
We have used the fact that, after absorption of a laser photon in the mode $\mathcal{E}_\mu$
and spontaneous emission of a photon of wavevector $k_L \mathbf{n}$, where $k_0$ was identified
to $k_L$, the initial center of mass atomic state of zero Bloch vector acquires a Bloch
vector $\mathbf{q}=k_L \mathbf{e}_\mu - k_L \mathbf{n}$.
The quantity $C$ results from the evaluation of the trace over the internal atomic variables
performed in the decoupled basis, where the dipole operator $\mathcal{D}$ has simple matrix elements
on $1/2\to 1/2$ and $1/2\to 3/2$ transitions. It only depends on the atom-laser detunings
$\delta_J$:
\be
\label{eq:C}
C = \frac{3}{2}\ \frac{1+2\delta_{\frac{3}{2}}/\delta_{\frac{1}{2}}}
{2+{\delta_{\frac{3}{2}}^2/\delta_{\frac{1}{2}}^2}}.
\ee
In principle, $C$ can range between $-3/4$ and $3/2$. In the useful case of detunings
much larger than the fine structure $\omega_{3/2}-\omega_{1/2}$, one has $\delta_{3/2}\approx
\delta_{1/2}$ and $C$ approaches $3/2$.

\begin{figure}[htb]
\centering{\includegraphics[width=8cm,clip=]{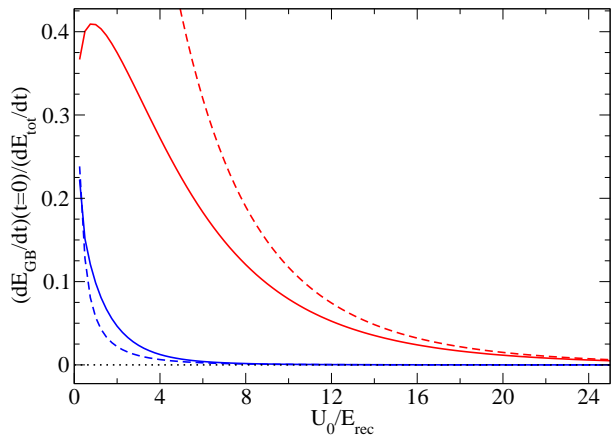}}
\caption{Initial ground band heating rate $dE_{\rm GB}/dt$
for an atom prepared in the ground Bloch state of the optical lattice,
for a red lattice (red lines) and for a blue lattice (blue lines),
as a function of the lattice depth in recoil units, $U_0/E_{\rm rec}$.
Here the coefficient $C=3/2$, and the lattice is isotropic:
$U_{0,\mu}=U_0$ for all spatial directions $\mu=x,y,z$ in Eq.(\ref{eq:Ured}) for a red lattice
and in Eq.(\ref{eq:Ublue}) for a blue lattice.
Solid lines: Numerical evaluation of (\ref{eq:degb_gen}). Dashed lines: Asymptotic expressions
(\ref{eq:asympt_red}) and (\ref{eq:asympt_blue}).
The ground band heating rate is expressed in units of the total heating rate (\ref{eq:de3}).
\label{fig:heating_gb}}
\end{figure}

The expression (\ref{eq:degb_gen}) allows a straightforward numerical evaluation of the heating
rate. It is convenient to introduce the lattice depths $U_{0,\mu}\geq 0$  along each direction $\mu\in\{x,y,z\}$,
such that Eq.~(\ref{eq:basic}) reads
\bea
\label{eq:Ured}
U_{\rm red}(\rr) &=& -P_g \sum_\mu U_{0,\mu} \cos^2 k_L r_\mu \\
\label{eq:Ublue}
U_{\rm blue}(\rr) &=& P_g \sum_\mu U_{0,\mu} \sin^2 k_L r_\mu
\eea
for a red detuned and a blue detuned lattice, respectively.
In Fig.\ref{fig:heating_gb}, we plot the numerical results for the ground band heating rate
in an isotropic lattice $U_{0,\mu}=U_0$ $\forall \mu$, either red or blue detuned, in units 
of the total heating rate (\ref{eq:de3}), as a function of the lattice depth in units of the recoil
energy (\ref{eq:erec}),
taking $C=3/2$. Whereas the ground band heating rate is of the same order as the total heating rate
for shallow lattices $U_0\lesssim E_{\rm rec}$, it drops to much smaller values in the tight-binding limit
$U_0\gg E_{\rm rec}$, the main effect being that the ground band width drops exponentially fast
with $(U_0/E_{\rm rec})^{1/2}$  in that limit. 
Furthermore, it is observed in Fig.\ref{fig:heating_gb} that the ground band heating
rate drops faster for a blue detuned lattice than for a red detuned one.
This we now explain analytically, not restricting to an isotropic lattice. 

\noindent {\sl Lamb-Dicke regime:} We first rewrite the matrix element appearing in (\ref{eq:degb_gen})
in terms of the ground band Wannier function $\Phi_{\mathbf{0}}(\rr)$, assumed as usual
to be real and normalized to unity:
\begin{multline}
\label{eq:wannier}
\langle \mathbf{0}; k_L(\mathbf{e}_\mu-\mathbf{n}) |
e^{-ik_L \mathbf{n}\cdot\mathbf{r}} \mathcal{E}_\mu(\mathbf{r}) |\mathbf{0};\mathbf{0}\rangle =
\\
\sum_{\RR} \int d^3r \
\Phi_{\mathbf{0}} (\rr-\RR) e^{-i k_L \mathbf{n}\cdot\mathbf{r}}
\mathcal{E}_\mu(\rr) \Phi_0(\rr)
\end{multline}
where the sum is taken over the locations of the lattice potential minima  $\RR\in (\pi/k_L)\mathbb{Z}^3$.
In the large lattice depth limit $U_{0,\mu}/E_{\rm rec}\gg 1$, the ground band dispersion relation
may be approximated by the tight-binding expression
\be
\label{eq:tbe}
E_{\mathbf{0};\mathbf{q}}-E_{\mathbf{0};\mathbf{0}} = \sum_{\nu\in\{x,y,z\}} 2 t_\nu [1-\cos(q_\nu \pi/k_L)].
\ee
The sum (\ref{eq:wannier}), to zeroth order in the tunneling amplitudes $t_\nu\geq 0$,  
reduces to the $\RR=\mathbf{0}$ term. 
This term may be evaluated by a Lamb-Dicke expansion in powers of $k_L a_\mu$,
resulting from series expansions in powers of $k_L r$, {\sl e.g.}\
\be
 e^{-i k_L \mathbf{n}\cdot \rr} = 1 - i k_L \mathbf{n}\cdot \rr + O(k_L r)^2.
\ee
We have introduced the size 
\be
a_\mu=\left(\frac{\hbar}{2m\omega_\mu}\right)^{1/2} = \frac{1}{k_L} \left(\frac{E_{\rm rec}}{4 U_{0,\mu}}\right)^{1/4}
\ee
of the ground state of the harmonic oscillator
of angular oscillation frequency $\omega_\mu$
that approximates $U(\rr)$ around $\rr=\mathbf{0}$ along direction $\mu$.
Restricting to leading order in the Lamb-Dicke expansion, we can also
approximate $\Phi_0$ with the Gaussian wavefunction of the harmonic oscillator ground state.

\begin{table}[ht!]
\centering
\begin{tabular}{ |c|c|c|c|c|c|c|c|c|c|c|c|}
\cline{1-6}
& & & & & \\
Atom  & Wavelength $\lambda_L$ & $E_{\rm rec}$ & $\left(\frac{dE_{\rm tot}}{dt}\right)/t_{\nu}$ & $\left(\frac{dE_{\rm GB}}{dt}\right)/t_{\nu}$ 
& Ref.\\
 & &  &  &  &\\
\hline \hline
$^{\rm 87}$Rb & $850$~nm & $3.2~$kHz  & $6~$s$^{-1}$ & $1~$s$^{-1}$ & \cite{greiner2002a} \\
 \hline
$^{\rm 87}$Rb & $750$~nm & $4.1~$kHz & $12~$s$^{-1}$ & $2\times10^{-2}~$s$^{-1}$  & \\
 \hline \hline
 $^{\rm 133}$Cs  & $1064$~nm & $1.3~$kHz & $1~$s$^{-1}$ & $0.1~$s$^{-1}$  & \cite{gemelke2009a} \\
 \hline
 $^{\rm 133}$Cs  & $780$~nm & $2.5~$kHz & $2~$s$^{-1}$ & $5\times10^{-3}~$s$^{-1}$ & \\

\hline \hline

$^{\rm 87}$Rb &x axis: $765$~nm &  $3.9~$kHz  &  $8~$s$^{-1}$ & $10^{-2}~$s$^{-1}$ &  \\
\hline
& y+z axis:$844$~nm  &  $2.5~$kHz  &  $2~$s$^{-1}$ & $0.3~$s$^{-1}$  &  \\
\hline
& Total: &   & $10~$s$^{-1}$ & $0.3~$s$^{-1}$  &   \cite{trotzki2009a} \\

\cline{1-6}
\end{tabular}
\caption{Total and ground band heating rates for two different experiments with $^{87}$Rb and $^{133}$Cs, for blue and red detuned
isotropic optical lattices of depth per axis $U_{0,\mu}=10~E_{\rm R}$.
The heating rates are estimated using the formulas (\ref{eq:asympt_red}) and (\ref{eq:asympt_blue}) obtained in the Lamb-Dicke limit,
and normalized by the ground band tunnel energy $t_{\nu}\approx 2 \times 10^{-2} E_{\rm R}$ defined in (\ref{eq:tbe}). 
For these parameters, the coefficient $C$ in Eq.~(\ref{eq:C}) is within $10~$\% of the asymptotic value $3/2$. 
The corrections to the light shift potential $U$ discussed in subsection \ref{subsec:eom}
are very small, $\delta U_{\rm hf}\sim 10^{-4}-10^{-5}$, $\delta U_{\rm micro}\sim 10^{-6}-10^{-7}$, $\delta U_{\rm 1/\delta^2}\sim 10^{-3}$. 
The last two lines refer to the bichromatic lattice of \cite{trotzki2009a}.} \label{Table1}
\end{table}

\noindent {\sl Red lattice:} We use the expansion
$\mathcal{E}_\mu(\rr) = \mathcal{E}_{0,\mu}  + O(k_L r_\mu)^2$ so that
\be
\langle \mathbf{0}; k_L(\mathbf{e}_\mu-\mathbf{n}) |
e^{-ik_L \mathbf{n}\cdot\mathbf{r}} \mathcal{E}_\mu(\mathbf{r}) |\mathbf{0};\mathbf{0}\rangle
= \mathcal{E}_{0,\mu}[ 1 + O(k_La_{\rm ho})^2],
\ee
where $a_{\rm ho}$ is the largest of the $a_\nu$.
Upon insertion in (\ref{eq:degb_gen}), we face angular integrals that can be calculated in spherical
coordinates of axis $\nu$, {\sl e.g.}\
\be
\int \frac{d^2n}{4\pi} n_\nu^2 \cos(\pi n_\nu) = \int_{-1}^{1} \frac{du}{2} u^2 \cos(\pi u) = -\frac{2}{\pi^2}.
\ee
We obtain the deep lattice asymptotic expression at time $t=0$:
\begin{multline}
\label{eq:asympt_red}
\frac{d}{dt} E_{\rm GB}^{\rm red}= 
\Gamma \left(\sum_J \frac{d_J^2}{(\hbar\delta_J)^2}\right)
 \sum_{\mu,\nu} 2 t_\nu |\mathcal{E}_{0,\mu}|^2  \\
\times \left[1+\frac{C}{\pi^2} (1+\delta_{\mu\nu})\right],
\end{multline}
where $\delta_{\mu\nu}$ is the Kronecker symbol.
Expression (\ref{eq:asympt_red}) asymptotically matches 
the numerical results, see Fig.\ref{fig:heating_gb}.
We thus find that, for a red detuned lattice,  the ground band heating rate
is smaller than the total heating rate by a factor scaling as the tunneling amplitude
over the recoil energy. This reduction factor simply originates from the energy width
of the ground band.

\noindent {\sl Blue lattice:}
The laser field amplitudes vanish in $\rr=\mathbf{0}$ and have the expansion
$\mathcal{E}_\mu(\rr) = \mathcal{E}_{0,\mu} k_L r_\mu + O(k_L r_\mu)^3$
so that \footnote{The identity $\langle \Phi_0| r_\mu |\Phi_0\rangle =0$
is exact (it is not an artifact of the Gaussian approximation for $\Phi_0$)
and results from the parity of $|\Phi_0\rangle$.}
\begin{multline}
\langle \mathbf{0}; k_L(\mathbf{e}_\mu-\mathbf{n}) |
e^{-ik_L \mathbf{n}\cdot\mathbf{r}} \mathcal{E}_\mu(\mathbf{r}) |\mathbf{0};\mathbf{0}\rangle
= \\
\mathcal{E}_{0,\mu} [-i (k_L a_\mu)^2 + O(k_L a_{\rm ho})^4].
\end{multline}
Insertion in (\ref{eq:degb_gen}) and angular integration give at time $t=0$:
\begin{multline}
\label{eq:asympt_blue}
\frac{d}{dt} E_{\rm GB}^{\rm blue}=\Gamma \left(\sum_J \frac{d_J^2}{(\hbar\delta_J)^2}\right) 
\sum_{\mu,\nu} 2 t_\nu |\mathcal{E}_{0,\mu}|^2 (k_L a_\mu)^4\\
\times \left\{\left(1+\frac{C}{3}\right)\left(\frac{1}{3}-\frac{1}{\pi^2}\right)-
\left(\frac{1}{5}-\frac{9}{\pi^4}\right)C\right. \\
\left.-\delta_{\mu\nu} \left[\frac{1}{\pi^2}+\left(\frac{33}{\pi^4}-\frac{11}{3\pi^2}\right)C\right]
\right\}.
\end{multline}
This can be further simplified, since $|\mathcal{E}_{0,\mu}|^2 a_\mu^4$ is actually
independent on the amplitude of the standing wave $\mathcal{E}_\mu$ and hence on the direction
of space.
Expression (\ref{eq:asympt_blue}) asymptotically matches the numerical results, see
Fig.\ref{fig:heating_gb}.
We thus find that the blue detuned lattice has a ground band heating rate 
that is reduced as compared to the red one (\ref{eq:asympt_blue})
by factors $(k_L a_\mu)^4\propto E_{\rm rec}/U_{0,\mu}\ll 1$.
The first factor $(k_L a_{\mu})^2$ originates from the reduction of the fluorescence rate
as compared to the red lattice, due to the Lamb-Dicke effect: Absorption of a laser photon
in the standing wave $\mathcal{E}_{0,\mu}\sin k_L r_\mu$ 
brings the atom mainly in the first excited band, with a small transition amplitude
$\propto k_L a_\mu$ leading to a reduced fluorescence rate $\propto (k_L a_\mu)^2$.
The second factor $(k_L a_{\mu})^2$ originates from the branching ratio of spontaneous emission
from the internal state $e$ in the first excited band to the internal state $g$ in the ground band.

\noindent {\sl Limits of validity:} The ground band energy increases
linearly in time (with the rate that we have calculated)
only for short times. We can estimate the onset of non-linearity
by considering double fluorescence cycles (with the atom arriving in the ground band
after the second cycle). For a red detuned (respectively
blue detuned) lattice, the probability of such a double cycle over
the time interval $t$ is $\propto (\Gamma_{\rm fluo}^{\rm max} t)^2$
[respectively $\propto (\Gamma_{\rm fluo}^{\rm max} t)^2 (k_L a_{\rm ho})^4$],
corresponding to a ground band energy increase $\propto t_\nu$. The corresponding
mean increase of energy is negligible
as compared to the single cycle contribution $\Gamma_{\rm fluo}^{\rm max} t
t_\nu$ [respectively $\Gamma_{\rm fluo}^{\rm max} t t_\nu (k_L a_{\rm ho})^4$]
as long as $\Gamma_{\rm fluo}^{\rm max} t \ll 1$ in both cases.

As a final remark, we point out that the ground band heating rate
for a blue lattice, being much smaller in the tight binding limit
than the one for a red lattice with the same depth $U_0/E_{\rm rec}$,
is thus much more sensitive to any additional heating effects
not included in our theoretical model, in particular to collisions
between ground band and excited band atoms in the many-body case.

\section{Conclusion}
\label{sec:conclusion}

We have performed a full quantum calculation of single atom heating rates 
in a far-detuned, three-dimensional optical lattice, including explicitly a realistic atomic internal structure 
and the fact that the superimposed laser standing waves have different frequencies, 
as in real experiments.

First, we have calculated the total heating rate, that is the rate
of increase of the total mechanical energy of the atom.
Remarkably, we have found a universal expression, 
independent of the initial internal and external atomic state,
and simply equal to the product of the recoil energy $E_{\rm rec}$ and 
of the maximal fluorescence rate $\Gamma_{\rm fluo}^{\rm max}$
that may be realized in the lattice.
The total heating rate is thus
independent of the sign of the laser detuning (red or blue). 

This general feature is easy to understand in the limiting case
of a deep lattice for an atom initially in the ground energy band: The total
heating rate is then determined by rare photon scattering events 
transferring the
atom  to excited bands at a rate which is independent of the sign
of the detuning.
In the blue-detuned case, the atom most probably arrives in the first
excited band after a scattering event, which increases the energy
by the oscillation quantum $\hbar\omega_{\rm osc}$, that is
by much more than the recoil energy; this however takes place
at a rate $\propto \Gamma_{\rm fluo}^{\rm max} E_{\rm rec}/\hbar\omega_{\rm osc}\ll \Gamma_{\rm fluo}^{\rm max}$ 
because of the Lamb-Dicke effect. The product of the rate and of the energy change is indeed
$\propto \Gamma_{\rm fluo}^{\rm max} E_{\rm rec}$.
In the red-detuned case \cite{wolf2000a}, the fluorescence rate, of order 
$\Gamma_{\rm fluo}^{\rm max}$, is much larger;
the atom however mainly returns to the ground band 
after a scattering event, due to the Lamb-Dicke effect, 
which increases the energy by much less than the recoil energy;
with a branching ratio $\propto E_{\rm rec}/\hbar\omega_{\rm osc}$, the atom however
arrives in the first excited band, which increases the energy by $\hbar\omega_{\rm osc}$ 
and results in a heating rate again $\propto \Gamma_{\rm fluo}^{\rm max} E_{\rm rec}$.

Second, for an atom initially prepared in the lowest Bloch state
of the lattice (a minimal single-particle model for the 
many-body superfluid state realized in experiments),
we have calculated the initial ground band heating rate. 
This is the rate of energy increase 
due to processes where the atom returns to the ground band of the lattice 
after a photon scattering event. 
We have derived analytical expressions of this rate in the deep lattice limit.
They show that, in contrast to the total heating rate, the ground band heating rate strongly
depends on the laser detuning, and is strongly suppressed for blue
detuned lattices:  It is of order $\Gamma_{\rm fluo}^{\rm max} t_0$ for a red deep  lattice,
and of order $\Gamma_{\rm fluo}^{\rm max} t_0 (E_{\rm rec}/\hbar\omega_{\rm osc})^2$ for blue
deep lattice, where $t_0$ is the atomic tunneling amplitude between neighbouring
sites.

A recent experiment \cite{trotzki2009a} reported a measured heating
rate of $\left( dE/dt \right)_{\rm exp}/t_0 \approx 3.5 {\rm s}^{-1}$,
where the tunnelling amplitude $t_0$ was essentially independent of
the spatial direction. This heating rate
was obtained by using Quantum Monte Carlo simulations
at several temperatures to calibrate the experimental data:
The decrease of the visibility of the interference pattern 
observed after free expansion was recorded over time, 
and compared to a ``visibility-energy'' abacus obtained from the simulations. 

The heating rate measured this way is unsurprisingly smaller than
the total heating rate that we calculate, see Table~\ref{Table1}.
Indeed, the Quantum Monte Carlo simulations take into account the 
ground band only, and we expect the measured interference
pattern to depend mostly on the ground band atoms. 
Hence, the most pertinent rate to compare the experiment to
is the ground band heating rate.
The measured heating rate is significantly larger than the ground
band heating rate, though, see Table~\ref{Table1}.
To resolve this discrepancy, one would have to include effects
that we did not consider in this article: Firstly, heating due to
technical noise in the apparatus, which should be quantified
for a specific experiment, and secondly, the role of collisions
in redistributing part of the energy from the excited
to the ground band. To our knowledge, the latter, more fundamental
many-body problem, is still quite unexplored \cite{hung2009a} and
provides an interesting direction for future work.

We acknowledge useful discussions with I. Bloch, S. Trotzki, L. Pollet, 
N. Prokof'ev, B. Svistunov, M. Troyer, K. Mur, and J. Dalibard.
The authors are members of IFRAF. This work was supported by the 
DARPA project OLE.

\appendix
\section{Derivation of the  ground state master equation}
\label{app:deriv}

In this Appendix, we provide details about the derivation of the master equation used in the main text. 
We use an interaction picture with respect to 
the kinetic plus hyperfine Hamiltonian
$H_0=\frac{\mathbf{p}^2}{2m} + H_{\rm hf}$, where operators are modified as 
$\tilde{X}(t) = e^{iH_0t/\hbar} X(t) e^{-iH_0t/\hbar}$.
The Bloch equations read
\begin{multline}
\label{eq:eg}
\frac{d}{dt} \tilde{\sigma}_{e_{J} g} =
(i\delta_{J}-\frac{\Gamma}{2}) \tilde{\sigma}_{e_{J} g} \\
+\frac{1}{i\hbar} \Big[\left(\tilde{V}_{\rm AL}(t)\right)_{e_{J} g} \tilde{\sigma}_{gg} -
\sum_{J'} \tilde{\sigma}_{e_{J} e_{J'}} \left(\tilde{V}_{\rm AL}(t)\right)_{e_{J'} g}
\Big]
\end{multline}
\begin{multline}
\label{eq:ee}
\frac{d}{dt} \tilde{\sigma}_{e_{J} e_{J'}} = 
[i(\delta_{J}-\delta_{J'})-\Gamma] \tilde{\sigma}_{e_{J} e_{J'}}
\\
+ \frac{1}{i\hbar} \left[ \left(\tilde{V}_{\rm AL}(t)\right)_{e_{J} g} \tilde{\sigma}_{g e_{J'}}
-\tilde{\sigma}_{e_{J} g} \left(\tilde{V}_{\rm AL}(t)\right)_{g e_{J'}}
\right]
\end{multline}
\begin{multline}
\label{eq:gg}
\frac{d}{dt} \tilde{\sigma}_{gg} = 
\tilde{\mathcal{L}}[\tilde{\sigma}_{ee}]
+\frac{1}{i\hbar} \sum_{J} \left[
\left(\tilde{V}_{\rm AL}(t)\right)_{g e_{J}} \tilde{\sigma}_{e_{J} g} -
\mbox{h.c.}
\right].
\end{multline}
The feeding term of the ground state by spontaneous emission is given in Schr\"odinger picture by
(\ref{eq:feed}).

We perform the series of approximations discussed in the main text.
Integrating formally Eq.~(\ref{eq:eg}), after omission of $\tilde{\sigma}_{ee}$, neglecting $\mathbf{p}^2/2m$ as compared to
$\hbar\Gamma$, and neglecting a transient of duration $1/\Gamma$, we obtain the steady state optical coherence
\be
\label{eq:eg_force}
\tilde{\sigma}_{e_{J} g}(t) = \int_0^{+\infty} \frac{d\tau}{i\hbar} \, e^{(i\delta_{J}-\Gamma/2)\tau}
\left(\tilde{V}_{\rm AL}(t-\tau)\right)_{e_{J} g}\tilde{\sigma}_{gg}(t-\tau).
\ee
By repeated integrations by parts of the integral over $\tau$, integrating the exponential
factor, we get a formal expansion of $\tilde{\sigma}_{e_{J} g}(t)$ in powers of $1/(\delta_{J}+i\Gamma/2)$,
that we turn into a series in $1/\delta_{J}$.

\noindent {\sl To order $1/\delta$:} The optical coherences in Schr\"odinger picture are given by
\begin{equation}
\label{eq:eg1}
\sigma_{e_{J} g}^{(1)}(t) = \frac{1}{\hbar\delta_{J}} \left(V_{\rm AL}(t)\right)_{e_{J} g}
\sigma_{gg}(t).
\end{equation}
This is familiar for a constant atom-laser coupling. We have shown that it holds 
even for a time dependent coupling provided that Eq.~(\ref{eq:cond2}) holds.
Inserting this relation Eq.~(\ref{eq:eg1}) in Eq.~(\ref{eq:ee}) gives in steady state
$\sigma_{e_{J} e_{J'}}^{(1)}=0$. 
After insertion of Eq.~(\ref{eq:eg1}) in Eq.~(\ref{eq:gg}) we then find that
$\sigma_{gg}^{\rm diag}$ has a purely conservative evolution of Hamiltonian
$H_{\rm hf}^g + \frac{\mathbf{p}^2}{2m} + U^{(1)}(\mathbf{r},t)$
with the time-dependent light shift potential:
\begin{equation}
U^{(1)}(\mathbf{r},t) = \sum_{J} \frac{1}{\hbar \delta_{J}} \left(V_{\rm AL}(t)\right)_{g e_{J}}
\left(V_{\rm AL}(t)\right)_{e_{J} g}.
\end{equation}
As expected $U^{(1)}/\hbar$ is of order $\Omega^2/\delta$ so that $d\tilde{\sigma}_{gg}/dt$ is at most
of this order; in general, $U^{(1)}$ can induce Zeeman and even
hyperfine ground state coherences \footnote{Assume that only one of the laser amplitudes $\mathcal{E}_\mu$ is non zero.
Then only the $U^{(2)}$ contribution given in (\ref{eq:U2}) and the feeding term (\ref{eq:feed}) 
can induce hyperfine ground state coherences.  Since one has usually $\Gamma \leq \Delta_{\rm hf}^{e,g}$,
the contribution of $U^{(2)}$ dominates.}

\noindent{\sl To order $1/\delta^2$:} From Eq.~(\ref{eq:eg_force}) we obtain
\be
\label{eq:eg2}
\tilde{\sigma}_{e_{J} g}^{(2)}(t) = \frac{-i}{\hbar\delta_{J}^2} \left[\frac{\Gamma}{2}+ \frac{d}{dt}\right]
\left[\left(\tilde{V}_{\rm AL}(t)\right)_{e_{J} g}\tilde{\sigma}_{gg}(t)\right].
\ee
We shall neglect $d\tilde{\sigma}_{gg}/dt$
with respect to $\Gamma \tilde{\sigma}_{gg}$ as allowed by Eq.~(\ref{eq:cond_deplct}).
Inserting Eq.~(\ref{eq:eg2}) in Eq.~(\ref{eq:gg}) has two effects. 
First it induces a small modification of the light shift potential of order Eq.~(\ref{eq:du3}):
\begin{multline}
\label{eq:U2}
U^{(2)}= 
\sum_{J} 
\frac{1}{(\hbar\delta_{{J}})^2} \Big\{\left(V_{\rm AL}(t)\right)_{g e_{J}}
\Big([H_{\rm hf},V_{\rm AL}(t)]\Big)_{e_{J} g}
\\
-\frac{i\hbar}{2} \Big[\left(V_{\rm AL}(t)\right)_{g e_{J}} \left(\frac{d}{dt}V_{\rm AL}(t)\right)_{e_{J} g}
-\mbox{h.c.}\Big]
\Big\}.
\end{multline}
Second it induces a lossy evolution of $\sigma_{gg}^{\rm diag}$, $\frac{d}{dt}\sigma_{gg}\Big|_{\rm loss}
=-\frac{1}{2}\{W,\sigma_{gg}\}$, with
\be
\label{eq:W2}
W^{(2)}= \sum_{J} \frac{1}{(\hbar\delta_{J})^2}
\left[\Gamma+\frac{d}{dt}\right] \left(V_{\rm AL}(t)\right)_{g e_{J}} \left(V_{\rm AL}(t)\right)_{e_{J} g}.
\ee
The adiabatic elimination of the excited state matrix elements up to order $1/\delta^2$, 
remarkably leads to an expression similar to the usual two-level atom case, 
despite the presence of the hyperfine Hamiltonian $H_{\rm hf}^{e_{J}}$ and
the time dependence of $V_{\rm AL}$:
\be
\label{eq:eef}
\sigma_{e_{J} e_{J'}}^{(2)} =
\frac{\left(V_{\rm AL}(t)\right)_{e_{J} g} \sigma_{gg}^{\rm diag}
\left(V_{\rm AL}(t)\right)_{g e_{J'}}}{\hbar^2\delta_{J}\delta_{J'}} + 
O\left[\left(\frac{\Omega}{\delta}\right)^2\frac{\Omega^2}{\delta\Gamma}\right].
\ee
This may me checked by direct insertion of Eq.~(\ref{eq:eg2}) in Eq.~(\ref{eq:ee}), the $O[\ldots]$
term coming from neglecting $d\tilde{\sigma}_{gg}/dt$ as allowed by Eq.~(\ref{eq:cond_deplct}).
When inserted in Eq.~(\ref{eq:gg}) this provides a closed equation for $\sigma_{gg}^{\rm diag}$.

\noindent{\sl Temporal average:} Under the conditions discussed in the main text,
we now average out the rapidly oscillating terms in the ground state master equation.
The lattice potential $U^{(1)}(\rr,t)$ to order $1/\delta$ contains
a zero frequency part, which is scalar according to Eq.~(\ref{eq:scalarite}) and
is called $U$ in Eq.~(\ref{eq:basic}).
It also contains oscillating contributions $U^{(1)}_{\mu \nu}(\rr) e^{-i\Delta_{\mu\nu}t}$,
with $\Delta_{\mu\nu}=\Delta_\mu-\Delta_\nu\neq 0$, $\mu\ne\nu$, supposed to be pairwise distinct, 
as it is the case in \cite{trotzki2009a},
and 
\be
\label{eq:Umunu}
U_{\mu\nu}^{(1)}(\rr)=
\sum_{J} \frac{(D_\nu)_{ge_{J}} \mathcal{E}^*_\nu(\rr)  (D_\mu)_{e_{J} g}\mathcal{E}_\mu(\rr)}{\hbar\delta_J}
\ee
These contributions are in general not
scalar. 
It is convenient to split $U_{\mu\nu}^{(1)}$ in (i) an off-diagonal part,
inducing coherences between the two hyperfine ground states, that is a raising (resp.\ lowering)
part $(U_{\mu\nu}^{(1)})_\pm$ coupling $g_1$ to $g_2$ (resp. $g_2$ to $g_1$),  and (ii) a diagonal part
$(U_{\mu\nu}^{(1)})_{\rm diag}$ coupling $g_1$ to $g_1$ and $g_2$ to $g_2$. In the interaction picture, the diagonal
parts correspond to terms modulated at frequencies $\Delta_{\mu\nu}$ in $\tilde{U}^{(1)}$,
and the off-diagonal parts to terms modulated at frequencies $\Delta_{\mu\nu}\mp\Delta_{\rm hf}^{g}$.
They will induce rapidly oscillating components of the density operator $\tilde{\sigma}_{gg}$ at those frequencies,
these components
being very small here since $\hbar\Omega^2/\delta \ll |\Delta_{\mu\nu}|, \Delta_{\rm hf}^{g}$.
In the present regime $\Delta_{\rm hf}^{g}\gg \Delta_{\rm mod}$, we can treat separately
the effect of the off-diagonal and diagonal parts of $U^{(1)}$.

The off-diagonal part of $U^{(1)}(t)$ induces a Rabi coupling $\approx \hbar\Omega^2/\delta$ between $g_1$ and $g_2$ much
smaller than their energy splitting $\hbar\Delta_{\rm hf}^g$, with a very slow time variation
at the scale of $1/\Delta_{\rm hf}^{g}$. Hence we treat it directly to second order in usual perturbation
theory for a fixed time $t$ and then average the result over time. This leads to an effective light shift potential
acting within each hyperfine ground state:
\be
\label{eq:1I}
U_{\rm eff,I}^{(1)} =  -\sum_{\mu\neq \nu} \frac{(U_{\nu\mu}^{(1)})_- (U_{\mu\nu}^{(1)})_+ 
- (U_{\nu\mu}^{(1)})_+ (U_{\mu\nu}^{(1)})_-}{\hbar \Delta_{\rm hf}^g}.
\ee
This corresponds to Eq.~(\ref{eq:du1}) with a different notation.
Having eliminated the hyperfine coherences, we consider in what follows
$\bar{\sigma}_{gg}^{\rm diag}$.

The diagonal part of $U^{(1)}_{\mu \nu}$ couples the zero frequency
component $\bar{\sigma}_{gg}^{\rm diag}$ to rapidly modulated components
$\sigma_{gg}^{\mu\nu}(t)e^{-i\Delta_{\mu\nu}t}$; adiabatic elimination gives
the slowly evolving amplitudes
\be
\sigma_{gg}^{\mu\nu}(t)=\frac{[(U^{(1)}_{\mu\nu})_{\rm diag},\bar{\sigma}_{gg}^{\rm diag}]}{\hbar\Delta_{\mu\nu}}.
\ee
They are small according to Eq.~(\ref{eq:cond_dmod}).
Including the coupling of $\bar{\sigma}_{gg}^{\rm diag}$ to
$\sigma_{gg}^{\nu\mu}$ by $(U_{\mu\nu}^{(1)})_{\rm diag}$ gives the {\sl a priori} leading contribution to the 
effective potential induced by the diagonal micromotion:
\be
\label{eq:1II}
U_{\rm eff,II}^{(1)} = \sum_{\mu\neq\nu} \frac{[(U_{\mu\nu}^{(1)})_{\rm diag}^\dagger,
(U_{\mu\nu}^{(1)})_{\rm diag}]}{2\hbar\Delta_{\mu\nu}}+\ldots
\ee
Usually, the micromotion is studied for a spinless particle, in which case the commutator in Eq.~(\ref{eq:1II}) 
automatically vanishes and the first non-zero correction scales as $1/\Delta_{\rm mod}^2$ \cite{rahav2003a,ridinger2009a}.
Here, the atom has a non-zero spin.
From a calculation in the decoupled basis we find \cite{dupontroc1972a}
$U_{\mu\nu}^{(1)}=\mathcal{E}_\mu \mathcal{E}_\nu^* (2id^2/3\hbar^2)\, \mathbf{J}\cdot(\mathbf{e}_\mu\times
\mathbf{e}_\nu)[\delta_{3/2}^{-1}-\delta_{1/2}^{-1}]$. The commutator in Eq.~(\ref{eq:1II}) thus 
vanishes also in our spinorial case for the field Eq.~(\ref{eq:champ}). Going to next order in $1/\Delta_{\rm mod}$ with the
formalism of \cite{rahav2003a} extended to the case with a spin we finally find a non-zero contribution
\be
\label{eq:1IIf}
U_{\rm eff,II}^{(1)} =  \sum_{\mu\neq\nu}
\frac{\mbox{\small grad}\, (U_{\mu\nu}^{(1)})_{\rm diag}^\dagger\cdot
\mbox{\small grad}\, (U_{\mu\nu}^{(1)})_{\rm diag}}{2m\Delta_{\mu\nu}^2}.
\ee
This corresponds to Eq.~(\ref{eq:du2}) with a different notation.
In appendix \ref{app:micro} we present an alternative derivation of this result, based on the dressed atom
approach, and showing that there is no other micromotion term of order $1/\Delta_{\rm mod}^2$.
Summing both contributions Eqs.(\ref{eq:1I},\ref{eq:1IIf}), 
we get the effective time-independent correction to the light shift potential for the theory of order $1/\delta$,
integrating out the effect of hyperfine couplings and diagonal atomic micromotions.

To eliminate the rapidly varying temporal components in the $1/\delta^2$ terms of $d\sigma_{gg}/dt$,
we simply take the temporal averages, neglecting micromotion corrections at this order, as allowed in particular
by Eq.~(\ref{eq:cond_deplct}). We also project out the operators inducing ground state hyperfine
coherences.
The $1/\delta^2$ correction $\bar{U}^{(2)}_{\rm diag}$ to the light shift is obtained by projecting
out the hyperfine coherences of Eq.~(\ref{eq:U2}) and taking the temporal average. Its explicit expression is
not required here. The complete expression for $\delta U$ in Eq.~(\ref{eq:ggf}) is
\be
\delta U\equiv U^{(1)}_{\rm eff,I}+U^{(1)}_{\rm eff,II}+\bar{U}^{(2)}_{\rm diag}.
\ee
The time average of Eq.~(\ref{eq:W2}) and the use of the scalarity relation
Eq.~(\ref{eq:scalarite}) gives the lossy part of Eq.~(\ref{eq:ggf}).
The time average of Eq.~(\ref{eq:eef}) gives Eq.~(\ref{eq:eeb}) with a different notation.

\section{A derivation of the micromotion potential based on the dressed atom picture}
\label{app:micro}

We propose here a derivation of the micromotion potential alternative to \cite{rahav2003a}
and including the atomic spin.
The idea is to use the dressed atom approach \cite{livre_CCT} to eliminate the time-dependence
of the Hamiltonian, and to use standard time-independent effective Hamiltonian theory \cite{livre_CCT}.
The laser field is then assumed to be initially in a Fock state with huge occupation numbers
$n_\mu$ in each mode $\mathcal{E}_\mu$ of frequency $\omega_L + \Delta_\mu$.
One may then neglect the dependence of the atom-laser coupling with the photon number,
replacing there the photon annihilation operators $a_\mu$ with $n_\mu^{1/2} A_\mu$,
where the phase operator $A_\mu=(a_\mu a_\mu^\dagger)^{-1/2} a_\mu$ 
has unit matrix elements in the Fock basis \cite{operateur_phase}.
The resulting Hamiltonian is $H=H_0+V$ with
\bea
H_0 &=& h_0 + \sum_{\mu} \hbar(\omega_L + \Delta_\mu) (a_\mu^\dagger a_\mu-n_\mu) \\
V &=& \sum_{\mu\neq\nu} \left(U^{(1)}_{\mu\nu}\right)_{\rm diag} A_\nu^\dagger A_\mu
\eea
where $h_0=\frac{\mathbf{p}^2}{2m} + \bar{U}$ is a purely atomic Hamiltonian,
$U^{(1)}_{\mu\nu}$ is given by (\ref{eq:Umunu}) and $\bar{U}$ is the time independent part of $U^{(1)}$, called $U$ in
(\ref{eq:basic}).
It is convenient to set  $(U^{(1)}_{\mu\nu})_{\rm diag} = \mathcal{E}_\nu^* \mathcal{E}_\mu T_{\mu \nu}$ where the operator $T_{\mu\nu}$
does not depend on the atomic position and obeys
\be
T_{\mu\nu} ^\dagger = -T_{\mu\nu} = T_{\nu \mu}.
\ee

Since the field is initially in the Fock state $|(n_\mu)_{\mu\in\{x,y,z\}}\rangle$, we introduce the orthogonal projector
$P$ on that Fock state, and $Q=1-P$ the supplementary projector. The effective Hamiltonian $H_{\rm eff}(E)$
inside the subspace over which $P$ projects is exactly given by
\be
\label{eq:heff}
H_{\rm eff}(E) = h_0 P + P V Q \frac{Q}{EQ-QHQ} QVP,
\ee
where we used $PVP=0$ resulting from the fact that $V$ does not conserve the number of photons in each mode.
At this stage, $E$ is arbitrary but much smaller than $\hbar \Delta_{\rm mod}$.
Since the denominator in (\ref{eq:heff}) is of order $\hbar \Delta_{\rm mod} \gg V$, because of the occurrence of the projector
$Q$, we may expand  (\ref{eq:heff}) in powers of $V$, restricting to terms of order up to $1/\Delta_{\rm mod}^2$:
\begin{multline}
\label{eq:expand}
H_{\rm eff}(E) = h_0 P + PVQ \frac{Q}{EQ- QH_0Q} QVP \\ + PVQ \frac{Q}{EQ-QH_0Q} QVQ \frac{Q}{EQ-QH_0Q} QVP  \\
+O(1/\Delta_{\rm mod}^3).
\end{multline}

We first focus on the terms of order $V^3$ in (\ref{eq:expand}). In the denominators, $\sum_{\mu} \hbar(\omega_L + \Delta_\mu) (a_\mu^\dagger a_\mu-n_\mu)$
gives contributions of order $\hbar\Delta_{\rm mod}$ that dominate over $E-h_0$. Expanding in powers of $(E-h_0)/\Delta_{\rm mod}$ then gives
\begin{multline}
\label{eq:e3}
H_{\rm eff}^{(3)}(E) = \sum'_{\alpha,\beta,\gamma} P |\mathcal{E}_\alpha|^2 |\mathcal{E}_\beta|^2 |\mathcal{E}_\gamma|^2
\left(
\frac{T_{\alpha\beta} T_{\gamma\alpha} T_{\beta\gamma}}{\hbar^2\Delta_{\alpha\beta} \Delta_{\gamma\beta}}\right. \\ +
\left.\frac{T_{\alpha\beta} T_{\beta\gamma} T_{\gamma\alpha}}{\hbar^2\Delta_{\alpha\beta} \Delta_{\alpha\gamma}} 
\right) P
+O(1/\Delta_{\rm mod}^3),
\end{multline}
where we recall that $\Delta_{\mu\nu}=\Delta_\mu-\Delta_\nu$, 
and the prime on $\sum$ means that the sum is taken over indices that are pairwise distinct. Since the modulation
frequencies $\Delta_\mu$ are incommensurate, a first action of $V$ on the Fock state $|(n_\mu)\rangle$,
{\sl e.g.}\ creating a photon in mode $\alpha$ and annihilating a photon in mode $\beta$, has to be compensated
in two steps, either annihilating $\alpha$ in the first step and creating $\beta$ in the second step,
or {\sl vice-versa}, hence the occurrence of two terms in (\ref{eq:e3}).
Exchanging the summation indices $\alpha$ and $\beta$ in the second term of (\ref{eq:e3}), and using
the antisymmetry of $T_{\alpha\beta}$ and $\Delta_{\alpha\beta}$ under the exchange $\alpha\leftrightarrow\beta$, a sum $1/\Delta_{\beta\gamma}+
1/\Delta_{\gamma\beta}=0$ appears, so that
\be
H_{\rm eff}^{(3)}(E) = O(1/\Delta_{\rm mod}^3),
\ee
and may be neglected.

We are left with the $V^2$ contribution to $H_{\rm eff}$:
\be
H_{\rm eff}^{(2)}(E) =  \sum_{\mu\neq\nu} P (U^{(1)}_{\mu\nu})_{\rm diag} \frac{1}{E-h_0- \hbar\Delta_{\mu\nu}} (U^{(1)}_{\nu\mu})_{\rm diag} P.
\ee
We expand this expression up to order $1/\Delta_{\rm mod}^2$:
\begin{multline}
H_{\rm eff}^{(2)}(E) =\sum_{\mu\neq\nu} P |\mathcal{E}_\mu|^2|\mathcal{E}_\nu|^2 T_{\mu\nu} \frac{1}{-\hbar\Delta_{\mu\nu}} T_{\nu\mu} P
\\
+  \sum_{\mu\neq\nu} P (U^{(1)}_{\mu\nu})_{\rm diag} (h_0-E) (U^{(1)}_{\nu\mu})_{\rm diag} P \frac{1}{(\hbar\Delta_{\mu\nu})^2} \\
+O(1/\Delta_{\rm mod}^3).
\end{multline}
The first contribution, of order $1/\Delta_{\rm mod}$, corresponds to the effective potential given in (\ref{eq:1II}) and vanishes, as already noted
in Appendix \ref{app:deriv},  since $T_{\mu\nu}$ and $\Delta_{\mu\nu}$
are antisymmetric functions of $\mu$ and $\nu$. The second contribution does not vanish. Since it is of order $1/\Delta_{\rm mod}^2$ already,
we treat it in perturbation theory: An unperturbed eigenstate $|\lambda_0\rangle$ of $h_0$ of
energy $E_0$ experiences an energy shift, calculated here up to order $(V/\Delta_{\rm mod})^2$, equal to
\be
E_2 = \langle \lambda_0| H_{\rm eff}^{(2)}(E_0) |\lambda_0\rangle.
\ee
Taking advantage of the fact that $\Delta_{\mu\nu}^2=\Delta_{\nu\mu}^2$,
we use the relation
\begin{multline}
\langle \lambda_0 | (U^{(1)}_{\mu\nu})_{\rm diag} (h_0-E_0) (U^{(1)}_{\nu\mu})_{\rm diag} |\lambda_0\rangle
+\mu\leftrightarrow\nu = \\
\langle \lambda_0 | [[(U^{(1)}_{\mu\nu})_{\rm diag},h_0],(U^{(1)}_{\nu\mu})_{\rm diag}]|\lambda_0\rangle.
\end{multline}
Since $\bar{U}$ is scalar, it commutes with $(U^{(1)}_{\mu\nu})_{\rm diag}$  and the kinetic energy operator $\mathbf{p}^2/(2m)$ is the only
one to contribute to the commutator. After an explicit calculation of the double commutator, we obtain
\be
E_2 =  \langle\lambda_0| U^{(1)}_{\rm eff,II}(\rr) |\lambda_0\rangle
\ee
where $U^{(1)}_{\rm eff,II}(\rr)$ is indeed given by (\ref{eq:1IIf}).
\bibliography{heatingBib}
\bibliographystyle{apsrev}

\end{document}